\documentclass[12pt]{article}
\begin{document}
\begin{center}
{\large \bf Comment on ``Solitonlike Solutions of the Grad-Shafranov Equation"}

\noindent
{\large  G. N. Throumoulopoulos$^\star$\footnote{
gthroum@artemis1.physics.uoi.gr}, K. Hizanidis$^\#$\footnote{
kyriakos@central.ntua.gr}, H. Tasso$\dag$\footnote{
het@ipp.mpg.de} \\
$^\star${\it University of Ioannina, \\ Association Euratom-Hellenic Republic,\\ Department of
Physics, GR 451 10 Ioannina,
Greece.}\\
$\#${\it National Technical University of Athens,\\ School of Electrical and Computer Engineering, \\ 9 Iroon Polytehniou Str., Zografou Campus, \\157 73 Athens, Greece} \\
$\dag$ {\it Max-Planck-Institut f\"{u}r Plasmaphysik,\\
Euratom association,\\ D-85748 Garching bei M\"{u}nchen, Germany}}
\end{center}
\begin{center}
{\large \bf Abstract}
\end{center}
In the above entitled  recent publication  by Giovanni Lapenta [Phys. Rev. Lett. {\bf 90}, 135005 (2003) ]  it is claimed construction of a new class
of solitonlike solutions for the Grad-Shafranov equation in plane geometry.
It is proved here that, because of the mathematically erroneous choice 
$\nabla p= |\Psi|^2 \Psi \nabla \Psi$
 for an analytic continuation of the poloidal magnetic flux-function $\Psi$ in the
complex plane ($p$ is the pressure), the cubic Schr\"odinger equation considered by the author is irrelevant to the equilibrium problem and the Grad-Shafranov equation.
\newpage

In a recent publication \cite{La}, the author claims derivation  of 
a new class of solitonlike solutions for the Grad-Shafranov equation in plane geometry.
The equilibrium equations considered are (Eqs. (5) [1])  
\begin{eqnarray}
\left(\nabla \Psi \times \nabla B_z\right)\cdot \hat{{\bf z}}&=&0 \nonumber \\
\nabla p + \nabla^2 \Psi \nabla \Psi + B_z\nabla B_z&=&0,
                                                        \label{1}
\end{eqnarray}
where $p(\Psi)$ and $B_z(\Psi)$ are the pressure and the $z$- component
of the magnetic field, respectively. The following forms of the free functions
$p$ and $B_z$ are then chosen (Eq. (13) [1])
\begin{equation}
B_z\nabla B_z = \alpha_0^2 \Psi\nabla \Psi
                                                   \label{2}
\end{equation}
\begin{equation}
\nabla p = \alpha_0^2 |\Psi|^2 \Psi \nabla \Psi,
                                                   \label{3}
\end{equation}
and (\ref{1}) is extended  in the complex plane, 
thus leading to the cubic Schr\"odinger equation (Eq. (14) [1])
\begin{equation}
\frac{\partial^2 \Psi}{\partial x^2} 
+ \frac{\partial^2 \Psi}{\partial y^2} = -\alpha_0^2(1+ | \Psi|^2) \Psi.
                                                    \label{4}
\end{equation}
A solitonlike solution of (\ref{4}) is (Eq. (15) [1])
\begin{equation}
\Psi(x,y)=\Psi_p sech (x/L)e^{-j(\alpha_0 +1/2\alpha_0y_0^2)y}.
                                                      \label{5}
\end{equation}

For a complex function $\Psi$, however, the rhs of (\ref{3}) becomes 
a non holomorphic function, i.e. owing to the fact that $|\Psi|$
does not have derivative, the term 
$\alpha_0^2|\Psi|^2\Psi\nabla \Psi$ 
can not be a function gradient as the lhs
of (\ref{3})   requires. An explicit proof follows. Taking   the curl of
(3)   yields 
$\nabla|\Psi|\times\nabla \Psi =0$,
implying   that  $\Psi$ depends only on $|\Psi|$:
\begin{equation}
 \Psi=f(|\Psi|).
                                               \label{5a}
\end{equation}
In order that the complex function $\Psi$ be analytic,  on 
 account of (\ref{5a}) and the Cauchy-Riemann conditions leads to 
$| \Psi|=$ constant and therefore $\Psi =$ constant.
Also, even without requiring analyticity  of $\Psi$,
by only considering its  polar form,
 $\Psi=|\Psi|\exp(j\Theta(x,y))$, Eq. (\ref{5a}) 
  implies that
\begin{equation}
\Theta=\Theta(|\Psi|).
                                                       \label{5b}
\end{equation}
 Solution  (\ref{5}), however,   is inconsistent
with (\ref{5b}) (otherwise it should hold that $x=x(y)$). Therefore, (4) is irrelevant to the Grad-Shafranov equation; as a matter
of fact the real part of (\ref{5}),
$$u(x,y)= \Psi_p sech (x/L)\cos \left\lbrack(\alpha_0 +1/2\alpha_0y_0^2)y\right\rbrack,$$ does not satisfy the respective Grad-Shafranov equation
\begin{equation}
\frac{\partial^2 u}{\partial x^2} 
+ \frac{\partial^2 u}{\partial y^2} = -\alpha_0^2(1+ u^2) u.
                                                    \label{6}
\end{equation}

A mathematically legitimate choice for
$\nabla p$, instead of (\ref{3}),   could be  
\begin{equation}
\nabla p = \alpha_0^2 \Psi^3 \nabla \Psi.    
                                                 \label{7}
\end{equation}
This leads to an equation of the form  (\ref{6}) for $\Psi$. Solving this equation in the complex plane, however, is a task not easier
than that for real $\Psi$.

In conclusion, because of the mathematically erroneous choice (3) for an
analytic continuation of $\Psi$, the cubic Schroedinger equation (4)
considered by
the author is irrelevant to the equilibrium problem (1) and to the
Grad-Shafranov equation. Despite this unlucky situation we consider the idea of the author as
appealing, and hope that it will be successful in the future if used in an
appropriate setting.

\begin{center}
{\large\bf Acknowledgements}
\end{center}

Part of this work was conducted during a visit of one of the
authors (G.N.T.) to  the Max-Planck Institut  f\"ur Plasmaphysik,
Garching. The hospitality of that Institute is greatly
appreciated. 

The present work was performed under the Contract of Association
ERB 5005 CT 99 0100 between the European Atomic Energy Community
and the Hellenic Republic.

\end{document}